\documentclass[pre,superscriptaddress,showkeys,showpacs,twocolumn]{revtex4-1}
\usepackage{graphicx}
\usepackage{latexsym}
\usepackage{amsmath}
\begin {document}
\title {Emergence of Social Structures  via Preferential Selection}
\author{Adam Lipowski}
\affiliation{Faculty of Physics, Adam Mickiewicz University, Pozna\'{n}, Poland}
\author{Dorota Lipowska}
\affiliation{Faculty of Modern Languages and Literature, Adam Mickiewicz University, Pozna\'{n}, Poland}
\author{Antonio Luis Ferreira}
\affiliation{Department of Physics and I3N, University of Aveiro, Portugal}
\begin {abstract}
We examine a weighted-network multi-agent model with preferential 
selection such that agents choose partners with the probability $p(w)$, where $w$ is the number of their past selections. When $p(w)$ increases 
sublinearly with the number of past selections ($p(w)\sim w^{\alpha}, \ \alpha<1$), agents 
develop a uniform preference for all other agents. At $\alpha=1$, this state looses stability and more complex structures form. For a 
superlinear increase ($\alpha>1$), strong heterogeneities emerge and 
agents make selections mainly within small and sometimes asymmetric clusters. 
Even in a few-agent case, formation of such clusters  resembles phase transitions with spontaneous symmetry breaking.
\end{abstract}

\maketitle
\section{Introduction}
The analysis of social networks is a rapidly growing research field, which draws an interdisciplinary interest of sociologists and psychologists, but also statisticians, computer scientists, and many others \cite{wasserman}. Such networks reflect the nature of social interactions, which are in fact responsible for many aspects of our life such as, e.g., opinion formation \cite{bahr}, disease spreading \cite{eubank}, business or friendship \cite{granovetter}. Since social networks develop in populations of interacting individuals, to understand formation and functioning of these intricate networks, one can use a statistical-mechanics approach, where global properties emerge from the behaviour of some basic building elements \cite{castellano2009}. 

Taking into account that social links are of various nature or strength, and moreover, they often vary in time, one might expect that an adequate description of social structures should be provided within the framework of evolving weighted networks. However, substantial complexity of such models hampers the analysis and their understanding is still rather poor \cite{yook2001}. One of the basic questions is whether social structures can be identified as certain phases or regimes in such weighted-network models. If so, one can ponder the nature of the transitions between such regimes and whether some asymmetries in a social structure (e.g., formation of leadership) could be related to symmetry breakings, which often accompany the transitions between phases. Such analysis would certainly link social networks with concepts of much wider applicability in studies of interacting systems. However, the complexity of many network models of social structure \cite{watts} suggests that gaining such a general and basic understanding might be very difficult. Let us also notice that evolving weighted networks are examples of complex networks \cite{barabasi}, and some insight in this field might possibly have broader implications.

The objective of the present paper is to examine a model of formation of some simple social structures. We identify regimes related to such structures and show that in some cases the transitions between these regimes resemble phase transitions with a spontaneous symmetry breaking, even in a tiny system of three individuals.  In our model, a continuous transition, which accompanies the emergence of an asymmetric leader in a small system, changes into a discontinuous transition in larger systems. Since the size of the system determines its  degeneracy, such behaviour suggests, despite apparent differences, a similarity to the  Potts model \cite{wu}. Moreover, at a certain transition point the behaviour of the model becomes more complex,  corresponding perhaps to real social structures.\\
\section{Model and Methods}
Let us consider $N$ agents, which repeatedly select each other with the probability that depends on the number of their previous selections.
The dynamics of the model is defined with the following rules: In each step, an agent (say $i$, where $i = 1,2,\ldots, N$) is selected with the uniform probability $1/N$. Using the roulette-wheel selection \cite{liproulette}, agent $i$ chooses another one (say $j\neq i$) with the probability proportional to $w_{i,j}^{\alpha}$, where $w_{i,j}(=w_{j,i})$ is the (time-dependent) weight associated with each pair of agents $(i,j)$,  and $\alpha>0$ is the preference exponent, which   relates weights to the selection probability. After selecting, the corresponding link is reinforced and its weight is increased by one ($w_{i,j}\rightarrow w_{i,j}+1$). Initially, unless specified otherwise, all weights are set to positive and less than unity random numbers. The reinforcement mechanism implemented in our model suggests a similarity to Polya urn model \cite{polya} and in the final part of our paper we will discuss this relation with some more details.

Some insight into the behaviour of our model can be obtained using arguments equivalent to the mean-field approximation (MFA). In particular, from the dynamical rules, one can deduce that the average evolution of weights (over independent runs) should approximately obey the following set of \mbox{$N(N-1)/2$} equations: 
\begin{equation}
\langle w_{i,j}\rangle_{t+1} = \langle w_{i,j}\rangle_{t}+N^{-1} \langle w_{i,j}\rangle_t^{\alpha}\left(w_i^{-1}+w_j^{-1}\right) 
\label{eq-w}
\end{equation}
where $t$ is the number of steps and $w_i=\sum_{k\neq i} \langle w_{i,k}\rangle_t^{\alpha}$.
Factors that contribute to the increase of $\langle w_{i,j}\rangle_t$ are actually of the form $\langle w_{i,j}^{\alpha}/ \sum_{k\neq i} w_{i,k}^{\alpha}\rangle$. Replacing the average of fractions with the fraction of averages and approximating $\langle w^{\alpha} \rangle \approx \langle w \rangle ^{\alpha}$, we obtain Eqs.~(\ref{eq-w}).
Although they are difficult to solve in general, it is plausible to assume that asymptotically (i.e., for large $t$),  $\langle w_{i,j}\rangle_t \approx a_{i,j}t$ and this transforms Eqs.~(\ref{eq-w}) into the following set of $N(N-1)/2$ nonlinear equations
\begin{equation}
a_{i,j} = N^{-1} a_{i,j}^{\alpha}\left(a_i^{-1}+a_j^{-1}\right) 
\label{eq-a}
\end{equation}
where $a_i=\sum_{k\neq i} a_{i,k}^{\alpha}$.  Let us notice that the  coefficients $a_{i,k}$ satisfy $\sum_{i<k} a_{i,k} = 1$, which follows from the fact that at each time step we increase the weight by one.  To supplement the analysis of the model and also to examine the validity of the above equations, we used Monte Carlo (MC) simulations implementing the dynamical rules of the model. Simulating the model for a large number of steps  $t$, we measured $\langle w_{i,j}\rangle /t$, which could be then compared with $a_{i,j}$.\\
\section{Results}
For $N=2$, the model has a trivial solution $w_{1,2}=w_{1,2}^0+t$ and hence $a_{1,2}=1$.  More interesting is the case $N=3$, where Eqs.~(\ref{eq-a}) can be explicitly written as follows
\begin{eqnarray}
a_{1,2} & = & \frac{1}{3}\left(\frac{a_{1,2}^{\alpha}}{a_{1,2}^{\alpha}+a_{2,3}^{\alpha}}+\frac{a_{1,2}^{\alpha}}{a_{1,2}^{\alpha}+a_{3,1}^{\alpha}}\right)  \nonumber\\
a_{2,3} & = & \frac{1}{3}\left(\frac{a_{2,3}^{\alpha}}{a_{2,3}^{\alpha}+a_{3,1}^{\alpha}}+\frac{a_{2,3}^{\alpha}}{a_{2,3}^{\alpha}+a_{1,2}^{\alpha}}\right)  \label{eq-a1} \\
a_{3,1} & = & \frac{1}{3}\left(\frac{a_{3,1}^{\alpha}}{a_{3,1}^{\alpha}+a_{1,2}^{\alpha}}+\frac{a_{3,1}^{\alpha}}{a_{3,1}^{\alpha}+a_{2,3}^{\alpha}}\right)  \nonumber
\end{eqnarray}
Some of the solutions of Eqs.~(\ref{eq-a1})  can be easily guessed.  In particular, one finds that (i) $a_{1,2}=a_{2,3}=a_{3,1}=1/3$ and (ii) $a_{1,2}=a_{2,3}=1/2,\ a_{3,1}=0$ (plus the other two permutations) are solutions of Eqs.~(\ref{eq-a1}) for any $\alpha$. Using numerical methods, one also finds for  $\alpha>3$ solutions of the form $a_{1,2}>a_{2,3}>a_{3,1}=0$ (plus the other five permutations). 
As we will see, solutions of this kind, where a single agent (say 1) is connected to all other agents and these are the only links in the system appear also for $N>3$. When a link with one agent (say 2) is strongest and all other $N-2$ links  are of the same strength, one can reduce eqs. (\ref{eq-a}) to a single equation
\begin{equation}
a_{1,2}=\frac{1}{N}\left[ 1+\frac{a_{1,2}^{\alpha}}{a_{1,2}^{\alpha}+(N-2)^{1-\alpha}( 1-a_{1,2} )^{\alpha}} \right] 
\label{mfa4}
\end{equation}
From the numerical solution of (\ref{mfa4}), we can calculate the remaining (nonzero) links $a_{1,3}=\ldots=a_{1,N}=\frac{1-a_{1,2}}{N-2}$.

\begin{figure}[]
\vspace{-0.5cm}
\includegraphics[width=\columnwidth]{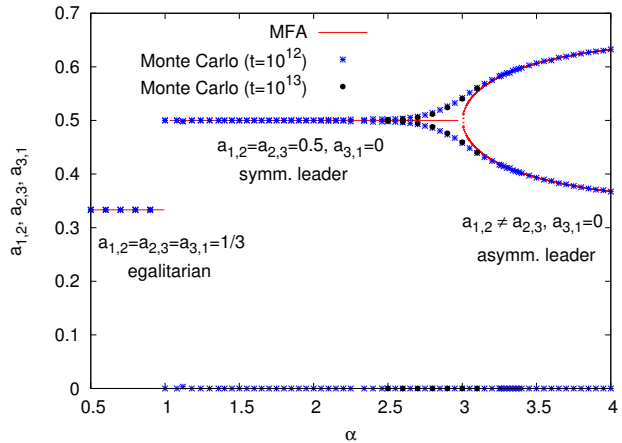}
\vspace{-0.8cm}
\caption{ (Color online) The asymptotic coefficients $a_{i,j}$ calculated for $N=3$  using the MFA (\ref{eq-a1}) and MC simulations. We started simulations from random initial configurations (and different for each $\alpha$).}
\label{3agents}
\vspace{-0.6cm}
\end{figure}

\begin{figure}
\centering
\includegraphics[width=0.7\columnwidth]{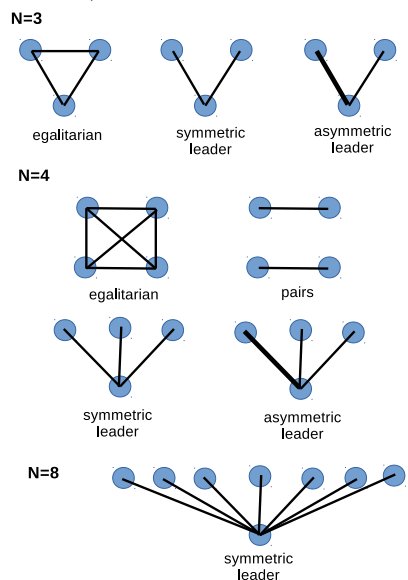}
\vspace{-6.mm}
\caption{ (Color online) Clusters that might emerge during the evolution of the model for $N=3$ and $N=4$.
To examine the stability of a symmetric leader for $N=8$ (Fig.~\ref{n8}) we used the initial configuration with the weights of the 7 indicated links set to unity ($w_{i,j}=1$) and the 21 remaining ones set to $10^{-2}$. 
}
\label{clusters}
\vspace{-0.6cm}
\end{figure}

However,  for a solution of Eqs.~(\ref{eq-a1}) to be a convergent solution of Eqs.~(\ref{eq-w}),  some stability conditions must be satisfied, namely that the modules of all  eigenvalues of the Jacobian are smaller than unity.  Omitting simple calculations, in Fig.~\ref{3agents} we present only the stable solutions of the MFA equations (\ref{eq-a1}). 
We interpret the solution $a_{1,2}=a_{2,3}=a_{3,1}=1/3$, which is stable for $\alpha<1$, as an egalitarian one, since all agents develop  equal preferences for each other (Fig.~\ref{clusters}). We call the solution of the type $a_{1,2}=a_{2,3}=1/2,\ a_{3,1}=0$ (which is stable for $1<\alpha<3$) a symmetric leader. In this case, agents 1 and 3 never select each other ($a_{3,1}=0$), and as a result agent 2 is selected more often than the other agents. Moreover, agent 2 has equal preferences towards agents 1 and 3 ($a_{1,2}=a_{2,3}=1/2$). For ${\alpha>3}$, this symmetry gets broken and the leader becomes asymmetric, i.e., preferentially selects one of the other agents (e.g., $a_{1,2}>a_{2,3}$). 

The comparison with extensive MC simulations ($t=10^{13}$) shows that despite simplifying assumptions, Eqs.~(\ref{eq-a1}) provide an accurate description of the model (Fig.~\ref{3agents}). Some discrepancy with simulations is seen only in the vicinity of  $\alpha=2.5$. Let us notice that at $\alpha=2.5$ we have a continuous transition between the regimes of symmetric and asymmetric leader, and the latter one has a spontaneously broken symmetry. In equilibrium statistical mechanics, transitions of this kind take place at critical points, the behaviour of which is known to deviate (in a sufficiently low dimension) from the mean-field description.
Strictly speaking, a broken symmetry in our model appears only in the limit $t\rightarrow\infty$. Only in such a case, there is a zero probability of a  jump that would change the 'vice-leader'.  The limit $t\rightarrow\infty$ is thus analogous to the thermodynamic limit that is needed to have a phase transition in equilibrium statistical mechanics. 

\begin{figure}
\centering
\includegraphics[width=\columnwidth]{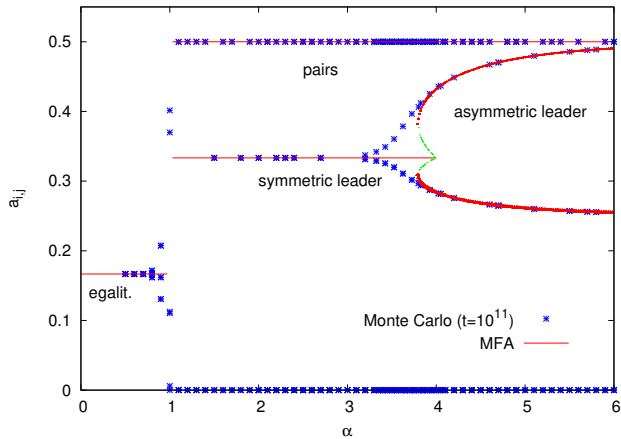}
\vspace{-7.mm}
\caption{ (Color online) The asymptotic coefficients $a_{i,j}$ calculated for $N=4$ using the MFA (\ref{eq-a}) and MC simulations. In the vicinity of $\alpha=4$ both 'symmetric' and 'asymmetric leader' solutions are stable (within MFA) and there is also an unstable branch of solutions (green dashed line).}
\label{4agents}
\vspace{-0.6cm}
\end{figure}

We made a similar analysis for $N=4$ (Fig.~\ref{4agents}). Our MC simulations show that in this case there is also an egalitarian solution as well as a symmetric and an asymmetric leader. Moreover, for ${\alpha}>1$ we often observe formation of clusters composed of two pairs (Fig.~\ref{clusters}).
Similar results are obtained from the solutions of   MFA eqs. (\ref{eq-a}), that for leader-type solutions simplify to  eq.~(\ref{mfa4}). Except the vicinity of $\alpha=4$, MC simulations are in a very good agreement with MFA  (Fig.~\ref{4agents}).  Let us notice that for $N=4$ the structure of the MFA solutions is different than that for $N=3$. In particular, upon reducing $\alpha$ the stable branch of an asymmetric-leader solution terminates  off the symmetric-leader solution ($\alpha\approx 3.9$). Moreover, the symmetric leader remains stable up to $\alpha=4$ and such behaviour resembles the hysteretic behaviour accompanying some discontinuous transitions.  However, the simulations for $N=4$ show that this is only an artefact of the mean-field approximation and the transition between a symmetric and an asymmetric leader is actually continuous and similar to the $N=3$ case.

\begin{figure}
\includegraphics[width=\columnwidth]{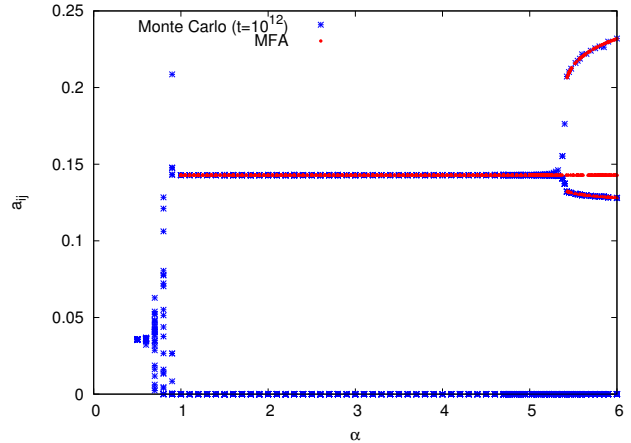}
\vspace{-0.7cm}
\caption{ (Color online) Asymptotic coefficients $a_{i,j}$ calculated for the $N=8$ system that starts as a 'symmetric leader' (Fig.~\ref{clusters}). Only a stable branch of MFA is plotted.}
\label{n8}
\vspace{-0.6cm}
\end{figure}

One might expect that a discontinuous transition should occur for larger $N$, however,  as we noticed starting from a random initial configuration, there is a very small probability of formation of clusters like a symmetric leader. A simple way out is to start from a configuration with a symmetric leader (Fig.~\ref{clusters}). Our numerical simulations for $N=8$ are in a good agreement with the MFA (eq.(\ref{mfa4}) for $N=8$) and confirm our expectations (Fig.~\ref{n8}). Let us notice that a symmetric leader is connected to $N-1$ agents and this determines the degeneracy of the system, which is removed in an asymmetric-leader regime. It is tempting to speculate that $N-1$ corresponds to the degeneracy of the ground state in systems such as the Potts model \cite{wu}. It is known that when such degeneracy is suffciently large, the symmetry-breaking transition in the Potts model becomes discontinuous. The above described transitions, which take place in our model even in a few-agent case and for which fluctuations in a nontrivial way modify the mean-field description, are novel and perhaps worth further studies.

\begin{figure}
\includegraphics[width=\columnwidth]{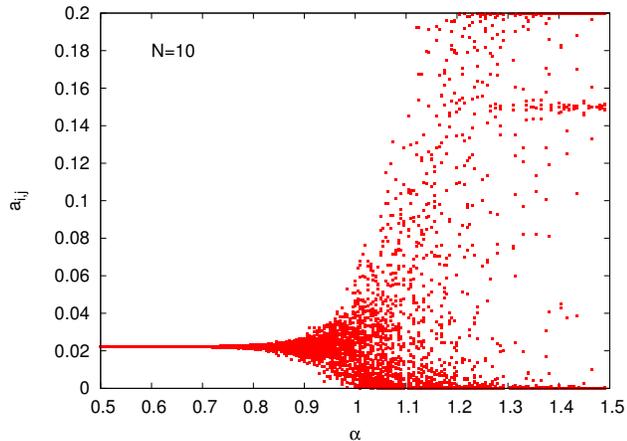}
\vspace{-0.7cm}
\caption{ (Color online) The asymptotic coefficients $a_{i,j}$ calculated for the $N=10$ system using MC simulations ($t=10^{10}$). Iteration of  MFA Eqs.~(\ref{eq-w}) gives similar results.}
\label{st9n10}
\vspace{-0.6cm}
\end{figure}

A similar analysis for larger $N$, especially within the MFA, is rather tedious because of a large number of links. 
Using MC simulations, we were able to get some insight into the overall behaviour of our model.
Omitting the analysis of various symmetry-breaking transitions for different clusters, which are possbile for larger $N$, we would like to show that an important change takes place at $\alpha=1$. Already for $N=10$ we can see that while for  $\alpha<1$ all  $a_{i,j}$ are positive, indicating a certain egalitarian type of solution,  for $\alpha>1$ a lot of $a_{i,j}$ vanish, which indicates formation of 
some (possibly small) clusters (Fig.~\ref{st9n10}). Let us notice, that the analysis we have made for (very) small $N$ is relevant also for larger $N$. In particular, for $N=10$ and $\alpha>1$ we observe for some runs decompositions into two pairs and two triples. For $\alpha<2.5$ these triples form symmetric leader solutions for which $a_{i,j}=\frac{1}{10}+\frac{1}{2}\cdot\frac{1}{10}=0.15$ and some accumulation around this value is seen in our MC data (Fig.~\ref{st9n10}). For $\alpha>2.5$ (not shown) these triples become asymetric leaders, in agreement with the analysis we made for $N=3$.

\begin{figure}
\includegraphics[width=\columnwidth]{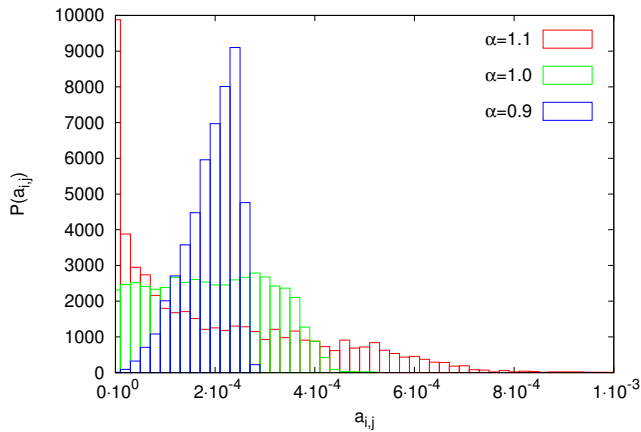}
\vspace{-0.9cm}
\caption{ (Color online) The distribution of $a_{i,j}$ calculated for $N=100$ using MC simulations ($t=10^{9}$). For $N=100$, there are 4950 links, hence in the egalitarian state we have $a_{i,j}=\frac{1}{4950}\approx 0.0002$, which is in a good agreement with the peak for $\alpha=0.9$.}
\label{a-distrib}
\vspace{-0.6cm}
\end{figure}

Clear indication of the transition at $\alpha=1$ is seen also for larger $N$.  Plotting the distribution of $a_{i,j}$ for $\alpha=0.9$, 1.0, and 1.1, we can see that, indeed, $\alpha=1$ marks the transition point between the two regimes of large and small concentrations of nearly-zero $a_{i,j}$ (Fig.~\ref{a-distrib}). Let us also notice that the distribution for $\alpha=1$ is nearly flat, which suggests that at this point the structure of clusters in the system is the most complex. It would be interesting to examine some other  characteristics for $\alpha=1$ and check whether our model might describe more realistic societies.

\begin{figure}
\includegraphics[width=7cm]{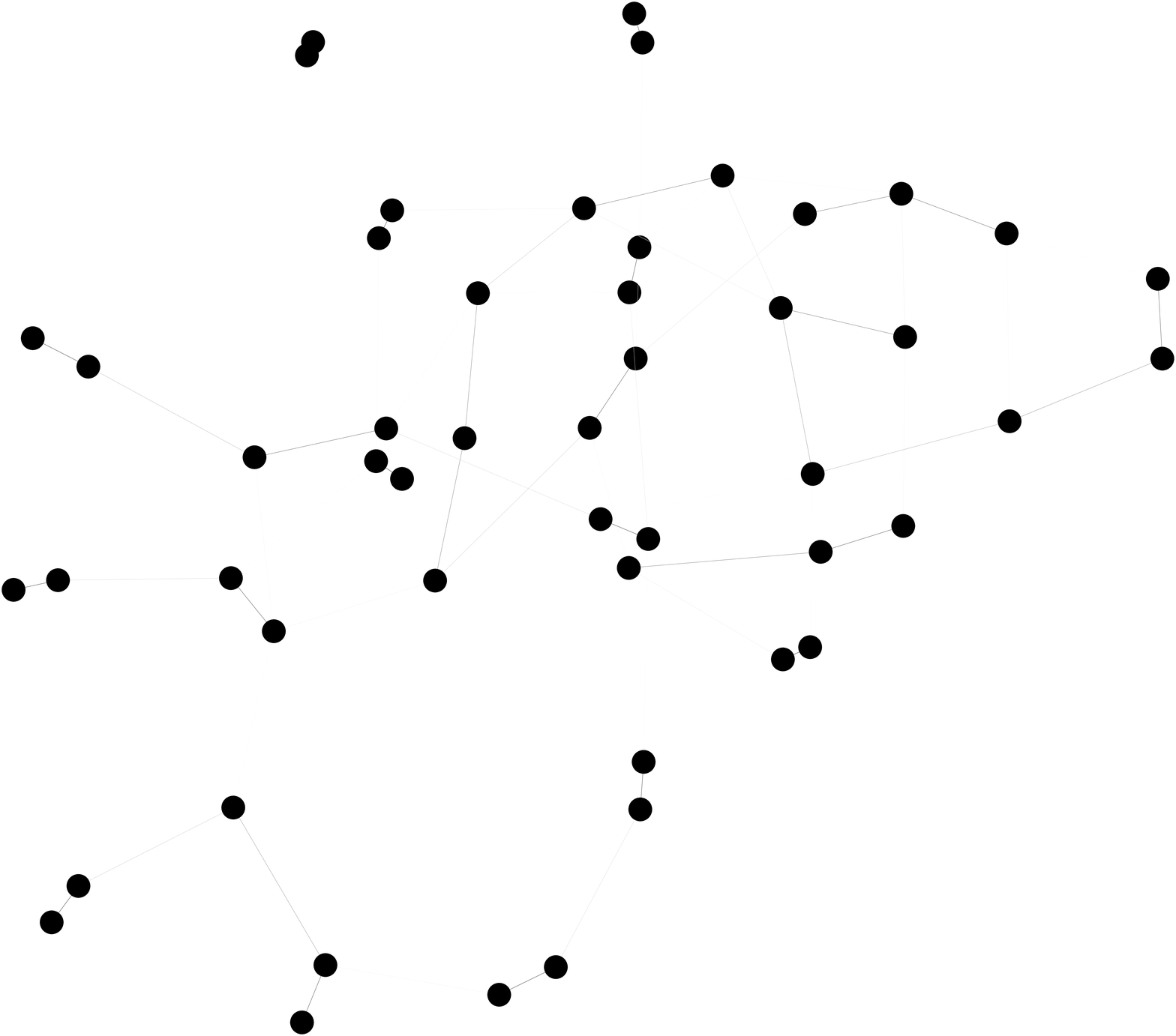}
\vspace{-0.0cm}
\caption{ The network structure of our model for $N=50$, $\alpha=1.4$
and after simulation time $t=10^9$. The weight of a link is represented
on the greyscale (the larger the weight the darker the link) and
approximately sets the length of a link (the larger the weight the
shorter the link).}
\label{graph14}
\vspace{-0.cm}
\end{figure}

\begin{figure}
\includegraphics[width=7cm]{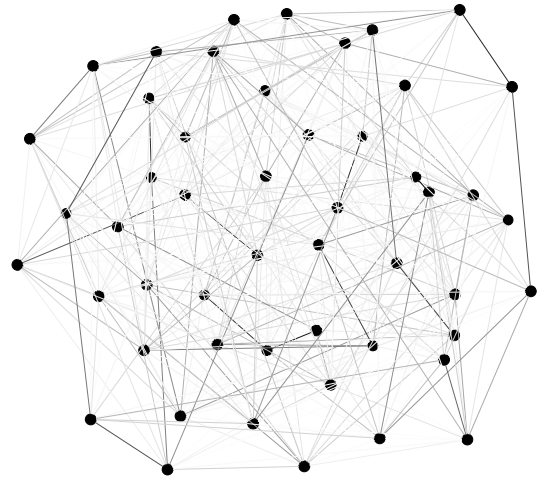}
\vspace{-0.0cm}
\caption{ The network structure of our model for $N=50$, $\alpha=1.2$
and after simulation time $t=10^9$. The rules for drawing the graph are the same here as in
Fig.~\ref{graph14}.}
\label{graph12}
\vspace{-0.cm}
\end{figure}

Drawing complex and in addition weighted networks is very difficult and
requires sophisticated techniques and software \cite{battista}.
Using Graphviz \cite{graphviz}, we were able to get some insight into the
structure of emerging clusters in the vicinity of the transition at
$\alpha=1$. One can notice that for larger $\alpha$ and in the limit
$t\rightarrow\infty$, the weighted network becomes sparse and only small
clusters form (Fig.~\ref{graph14}). For $\alpha$ closer to 1, the network
becomes more dense but still remains heterogeneous (Fig.~\ref{graph12}).

There are also some analytical arguments that $\alpha=1$ marks the transition point of our model for arbitrary $N$. Let us make a stability analysis of an egalitarian state, where  $a_{i,j}=D^{-1}$, and $D=\frac{N(N-1)}{2}$ is the number of links in the system. Elementary calculations show that $D\times D$ Jacobian $J$ for Eqs.~(\ref{eq-a}) has elements of the form
\begin{equation}
J_{k,l}=\alpha\left( \delta_{k,l}-\frac{1}{N-1} \right), \ \ k,l=1,2,\ldots, D
\label{jacobian}
\end{equation}
where $\delta_{k,l}$ is the Kronecker delta function. To find the spectrum of $J$, one can notice that it is a circulant matrix \cite{periodic} or one can use the decomposition $J=\alpha(I_D-\frac{1}{N-1}K_D)$, where $I_D$ and $K_D$ are the $D$-dimensional identity matrix and the matrix of ones \cite{ones}, respectively.
Hence,  $J$ has the following spectrum
\begin{equation}
\lambda_1=\lambda_2=\ldots = \lambda_{D-1}=\alpha, \ \lambda_D=\alpha(1-N/2)
\label{spectrum}
\end{equation} 
From the spectrum (\ref{spectrum}), one obtains that for $N=3$ and 4 the egalitarian state looses stability at $\alpha=1$. For larger $N$, one could expect at first sight that the egalitarian state looses stability at smaller $\alpha$ ($=\frac{2}{N-2}$) due to the last eigenvalue $\lambda_D$. However, the $D$-th eigenvector has the form $\frac{1}{\sqrt{D}}(1,1,\ldots,1)$ and thus any perturbation of the egalitarian state projected on this eigenvector will have a zero component (the dynamics of the MFA (\ref{eq-a}) keeps $\sum_{i<j} a_{i,j}$ constant). Consequently, $\lambda_D$  is irrelevant for the stability  of the MFA dynamics and the remaining eigenvalues indicate that the egalitarian state looses stability at $\alpha=1$. Our numerical results for $N=10$ show that the egalitarian state remains stable much beyond $\alpha=\frac{2}{10-2}=1/4$ and support such conclusion (Fig.~\ref{st9n10}). Even though the numerical simulations (Fig.~\ref{st9n10}, Fig.~\ref{a-distrib}) suggest that  for $\alpha<1$ the egalitarian state is not the only attractor of the dynamics, it might be a result of a slow convergence, especially in the vicinity of $\alpha=1$.\\
\section{Remarks and conclusions}
As we have already mentioned, our model might be considered as a certain version of Polya urn model \cite{polya}. In the original formulation of the urn model, one considers a collection of bins (urns) and balls that are added to a selected bin. Various forms of the probability of selection depending on the number of balls in a given bin were examined, including nonlinear ones, which are analogous to our $w^{\alpha}$.  In such a case, one finds \cite{nonlin} two regimes (i) with equal distribution of balls among bins ($\alpha<1$) and (ii) with a condensation of balls in a single bin ($\alpha>1$). Our model with egalitarian ($\alpha<1$) and clustered ($\alpha>1$) phases is clearly related to such nonlinear urn models. Actually, our model would be equivalent to the latter one if we directly selected a link instead of randomly selecting first an agent and only then a link (with preferences). For such link-oriented rules and for $\alpha>1$, we would  observe an  accumulation of weight on a certain (single) link. Instead, agent-oriented rules implemented in our model break the equivalence with the nonlinear urn model as they enforce some distribution of weights and formation of clusters. In the context of modeling of social structures, such agent-oriented rules seem to be more relevant. Let us also notice that for agent-oriented rules, the so-called exchangeability of probabilities fails and such models are much more  difficult for mathematical treatment \cite{skyrms,comm1}. Similarly to nonlinear (link-oriented) urn models, we expect that the transition in our model takes place at $\alpha=1$, but our arguments are based on the analysis of the mean-field approximation and more rigorous approach would be certainly desirable.

In conclusion, in the present paper we introduced a simple urn model of formation of social structures via a preferential-selection mechanism. Our model predicts the transition between two phases: (i) the egalitarian one, where agents select each other with equal probabilities and (ii) the phase, where agents select each other only within small clusters. Some of the clusters are asymmetric, which suggests that the model might describe more complex social structures with leaders or vice-leaders.  
Formation of asymmetric clusters for a few individuals proceeds via continuous symmetry-breaking transitions, which turn into discontinuous ones for larger systems, suggesting an intriguing similarity to the Potts model. It is tempting to think of some sociological implications of this result. In particular, we might expect that in a large group of people a vice-leader is unlikely to be weak (she/he emerges via a discontinuous jump with a large departure from the symmetric leadership), while in a small group ($N=3,\ldots,7$) a weak vice-leadership is more feasible. Sociology of small groups is an active research area \cite{smallgroup}, and we hope to check the relevance of our considerations.  The most complex structures form  in our model at $\alpha=1$, which is the transition point between the two phases. Although linearity is usually considered the simplest, in our model it offers the richest structures, which might suggest that complex societies also operate in (or close to) such a regime.

Acknowledgements: The research for this work was supported by NCN grant 2013/09/B/ST6/02277 (A.L.), NCN grant 2011/01/B/HS2/01293 (D.L.), and project PEst-C/CTM/LA0025/2013 from Funda\c c\~ao para a Ci\^encia e a Tecnologia (A.L.F.).



\begin{thebibliography}{}
\bibitem{comm1} For agent-oriented models one can restore exchangeability introducing dynamics with asymmetric links ($w_{i,j}\neq w_{j,i}$). In such a case the model effectively decomposes into $N$ link-oriented models \cite{skyrms}.
\end{thebibliography}
\end {document}